\documentclass{iau} 
\usepackage{graphicx}

\title[OH maser distributions in GC] 
{Ground-state OH maser distributions \\in the Galactic Centre region}

\author[Hai-Hua Qiao, Andrew J. Walsh, Zhi-Qiang Shen \& Joanne R. Dawson]   
{Hai-Hua Qiao$^{1,2,3,4}$
 Andrew J. Walsh$^2$ Zhi-Qiang Shen$^{1,3}$ \and Joanne R. Dawson$^5$}

\affiliation{$^1$Shanghai Astronomical Observatory, Chinese Academy of Sciences,\\ 80 Nandan Road, Shanghai, China, 200030 \\ email: {\tt haihua.qiao@student.curtin.edu.au} \\[\affilskip]
$^2$International Centre for Radio Astronomy Research, Curtin University,\\ GPO Box U1987, Perth WA 6845, Australia\\[\affilskip]
$^3${Key Laboratory of Radio Astronomy, Chinese Academy of Sciences, China}\\[\affilskip]
$^4${University of Chinese Academy of Sciences, 19A Yuquanlu, Beijing, China, 100049} \\[\affilskip]
$^5$Department of Physics and Astronomy and MQ Research Centre in Astronomy, Astrophysics and Astrophotonics, Macquarie University, NSW 2109, Australia}
\pubyear{2016}
\volume{322}  
\jname{The Multi-Messenger Astrophysics of the Galactic Centre}
\editors{Steve Longmore, Geoff Bicknell \& Roland Crocker, eds.}

\begin{document}

\maketitle

\begin{abstract}
Ground-state OH masers identified in the Southern Parkes Large-Area Survey in Hydroxyl were observed with the Australia Telescope Compact Array to obtain positions with high accuracy ($\sim$1\,arcsec). We classified these OH masers into evolved star OH maser sites, star formation OH maser sites, supernova remnant OH maser sites, planetary nebula OH maser sites and unknown maser sites using their accurate positions. Evolved star and star formation OH maser sites in the Galactic Centre region (between Galactic longitudes of $-5^{\circ}$ to $+5^{\circ}$ and Galactic latitudes of $-2^{\circ}$ and $+2^{\circ}$) were studied in detail to understand their distributions. 
\end{abstract}

\firstsection 
\section{Introduction}

Ground-state OH masers (i.e. 1612, 1665, 1667 and 1720\,MHz) are usually found towards the circumstellar envelopes of evolved giant and supergaint stars (e.g. Miras and OH/IR stars; \cite[Nguyen-Q-Rieu et al. 1979]{Nye1979}), regions of high-mass star formation (\cite[Reid \& Moran 1981]{RM1981}), supernova remnants (SNRs; \cite[Goss \& Robinson 1968]{GR1968}), centres of active galaxies (\cite[Baan et al. 1982]{Bae1982}), very rarely in planetary nebulae (PNe; \cite[Qiao et al. 2016]{Qie2016}) and comets (\cite[G{\'e}rard et al. 1998]{Gee1998}). 

SPLASH (the Southern Parkes Large-Area Survey in Hydroxyl) is an unbiased survey which observed all four ground-state OH transitions with the Australia Telescope National Facility Parkes 64-m telescope. SPLASH surveyed about 152 square degrees of the southern Galactic plane between Galactic longitudes of $332^{\circ}$ to $10^{\circ}$, crossing through the Galactic Centre, and Galactic latitudes of $-2^{\circ}$ and $+2^{\circ}$ (\cite[Dawson et al. 2014]{Dae2014}).
As a result, many sites with OH masers were detected. In order to identify these OH masers' astrophysical associations, we performed Australia Telescope Compact Array observations to obtain accurate positions for these SPLASH-detected OH maser emission. Each region was observed typically with five snapshot observations (duration 6 min each), which gave a typical on-source integration time of 30 min. The position accuracy was about 1 arcsec.
With accurate positions, we categorised these OH masers according to their astrophysical associations. We classified these OH masers into evolved star OH masers, star formation OH masers, SNR OH masers, PN OH masers and unknown OH masers.

This paper presents distribution of OH masers associated with evolved stars and star forming regions in the Galactic Centre region, between Galactic longitudes of $-5^{\circ}$ to $+5^{\circ}$ and Galactic latitudes of $-2^{\circ}$ and $+2^{\circ}$.

\section{Result}

\begin{figure}[h]
 \vspace*{-0.45 cm}
\begin{center}
 \includegraphics[width=3.2in]{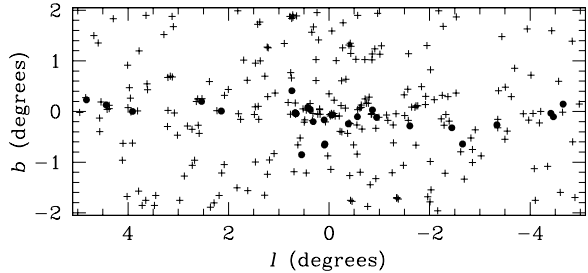} 
 \vspace*{-4.3 cm}
 \caption{OH maser distributions in the Galactic Centre region. Pluses are evolved star OH masers and filled circles are star formation OH masers.}
   \label{fig1}
\end{center}
\end{figure}

There are about 256 evolved star OH maser sites and about 31 star formation OH maser sites in the Galactic Centre region. Fig. \ref{fig1} shows their distributions. 

\section{Implications}

{\underline{\it Evolved star OH masers}}. Evolved star OH masers are widely distributed in Galactic latitude because they are either close to us, or they originate in the Galactic Bulge. We can see the density enhancement in the Galactic Bulge.

{\underline{\it Star formation OH masers}}. Star formation OH masers are tightly constrained to the Galactic Plane. There are more star formation OH masers in the negative Galactic latitude (21 out of 31). It is interesting to note that the distribution of star formation OH masers is evenly distributed with respect to Galactic longitude. These OH masers do not follow the dense gas concentrations that dominate at positive Galactic longitudes. Compared to the SPLASH pilot region ($334^{\circ} < l < 344^{\circ}$ and $-2^{\circ} < b < +2^{\circ}$; Qiao et al. in prep.; 63 star formation OH maser sites), there are less star formation OH maser sites in the Galactic Centre region, which contains more dense gas.
\\
\\
\textbf{Acknowledgements:} This work was supported in part by the Major Program of the National Natural Science Foundation of China (Grant No. 11590780, 11590784) and the Strategic Priority Research Program ``The Emergence of Cosmological Structures'' of the Chinese Academy of Sciences (Grant No. XDB09000000).

\end{document}